\documentclass[aps,pra,reprint,superscriptaddress,letterpaper,showpacs]{revtex4-1}
\usepackage[english]{babel}
\usepackage{ucs}
\usepackage[utf8x]{inputenc}
\usepackage{amsmath}
\usepackage{amsfonts}
\usepackage{amssymb}
\usepackage{graphicx}
\usepackage{bm}
\usepackage{bbm}
\usepackage{braket}

\newcommand{\col}{}

\begin{document}

\title{The Role of Multilevel Landau-Zener Interference in Extreme Harmonic Generation}
\author{J. Stehlik}
\affiliation{Department of Physics, Princeton University, Princeton, New Jersey 08544, USA}
\author{M. Z. Maialle}
\affiliation{Faculdade de Ci\^{e}ncias Aplicadas, Universidade Estadual de Campinas, 13484-350 Limeira, Sao Paulo, Brazil}
\author{M. H. Degani}
\affiliation{Faculdade de Ci\^{e}ncias Aplicadas, Universidade Estadual de Campinas, 13484-350 Limeira, Sao Paulo, Brazil}
\author{J. R. Petta}
\affiliation{Department of Physics, Princeton University, Princeton, New Jersey 08544, USA}
% 71.70.Ej - spin orbit in CM
% 73.63.Kv - electronic transport in QD
% 76.30.-v - Electron spin resonance
% 85.35.Gv - single electron devices

\begin{abstract}
Motivated by the observation of multiphoton electric dipole spin resonance processes in InAs nanowires, we theoretically study the transport dynamics of a periodically driven five-level system, modeling the level structure of a two-electron double quantum dot.  We show that the observed multiphoton resonances, which are dominant near interdot charge transitions, are due to multilevel Landau-Zener-St\"uckelberg-Majorana interference.  Here a third energy level serves as a shuttle that transfers population between the two resonant spin states.  By numerically integrating the master equation we replicate the main features observed in the experiments: multiphoton resonances (as large as 8 photons), a robust odd-even dependence, and oscillations in the electric dipole spin resonance signal as a function of energy level detuning.
\end{abstract}

\maketitle

\section{Introduction}

Harmonic generation occurs in a nonlinear system when driving at frequency $f$ results in a physical response of the system at multiples of the driving frequency, e.g.\ $2f$, $3f$, $4f$, and underpins nonlinear and quantum optics \cite{PhysRevLett.7.118,quantumoptics}.  Two-photon absorption can be observed in optically pumped systems at high powers \cite{PhysRevLett.9.453,PhysRevLett.7.229}.  Harmonic generation has also been observed in semiconductor systems that are driven with terahertz pulses \cite{Sherwin} and in electrically driven quantum dots \cite{NadjPergeSpectroscopy,HalfFreq1}.  Generally, multiphoton resonances are only observed at very high drive fields \cite{cohen}.  As a result, experimental observations are often limited to two-photon processes.

Multiphoton resonances were recently observed in electric dipole spin resonance (EDSR) in nanowire and planar quantum dots \cite{NadjPergeSpectroscopy,StehlikPRL,HalfFreq1}. Early experiments in these systems demonstrated  electric driving of single electron spins \cite{StefanNP}. These data were largely consistent with theoretical predictions, with an EDSR response observed when $h f = E_{\mathrm{z}i}$, where $E_{\mathrm{z}i}= g_i \mu_{\rm B} B$ is the Zeeman energy of the $i$-th dot, $h$ is Planck's constant, $f$ is the frequency of the electric driving field, $g_i$ is the electron $g$-factor of $i$-th dot, $\mu_{\rm B}$ is the Bohr magneton and $B$ is the applied magnetic field \cite{GolovachLoss,SpinOrbitQubit}.  However, more detailed investigations revealed that the multiphoton resonances were strongest when the double quantum dot (DQD) was driven near the interdot charge transition \cite{StehlikPRL}.  The EDSR harmonics, indexed by integer $n$, followed the resonance condition $n h f = E_{\mathrm{z}i}$ and showed a remarkable odd-even dependence, wherein the sign of the EDSR signal differed for odd and even $n$.  

In this paper we develop a full model of the DQD, building upon a three-level model of Danon and Rudner \cite{RudnerPRL2014}. We start by calculating the time-evolution of a simple five-level system, which captures the physics of a two-electron DQD. These simulations demonstrate that when the DQD is initialized in a spin-blocked state the system can make a Landau-Zener-St\"uckelberg-Majorana (LZSM) transition to an intermediate state, before making a final LZSM transition to a resonant unblocked state.  Thus the harmonics can be understood as being a multi-level LZSM effect  \cite{ShytovPRA}.

The EDSR resonances were observed in transport measurements. Therefore, to realistically model the experimental system, we add coupling to source-drain electrodes, decoherence, and charge noise \cite{TemchenkoPRB}. Our work extends the simple three-level model presented in Ref.~\cite{RudnerPRL2014} to a complete 5-level system that accurately describes a two-electron singlet-triplet qubit \cite{PettaSeminal}. The additional levels are found to contribute to the observed resonances and allow us to make a quantitative comparison with the experimental data.

The outline of the paper is as follows. In Sec.~\ref{sec:TLS} we describe the dynamics of a driven two-level system (TLS). The celebrated LZSM equation is introduced before showing that periodic driving gives rise to $n$-photon resonance conditions.  In Sec.~\ref{sec:DQD} we describe the singlet-triplet energy level diagram of a doubly occupied DQD, and show qualitatively how multilevel LZSM interference gives rise to harmonic generation. Finally in Sec.~\ref{sec:Results}, we add the effects of lead coupling and decoherence to our model. The calculated response is compared with the experimental results and shown to reproduce the key features observed in the data \cite{StehlikPRL}.

\section{Two-Level Landau-Zener-St\"uckelberg-Majorana Dynamics}
\label{sec:TLS}

\begin{figure}
\begin{center}
\includegraphics[width=\columnwidth]{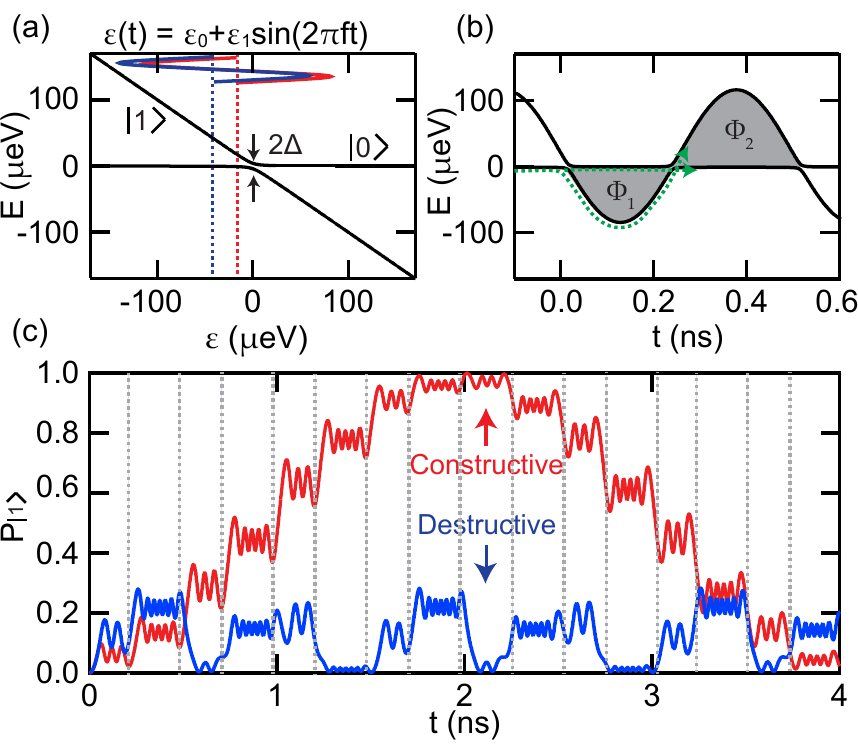}
\caption{\col  (a) Energy level diagram described by Eq.~\ref{Eq:TLS_H}.  Sinusoidal driving with $\epsilon(t) = \epsilon_0 + \epsilon_1 \sin(2\pi f t)$ causes repeated traversals of the anticrossing. (b) Energy levels as a function of time for sinusoidal driving. Each passage through the anticrossing results in some population transfer, analogous to a beam splitter. A phase $\Phi_{i}$ is accumulated between beam splitter events. (c) Population of the $\ket{1}$ state, $P_{\ket{1}}$, as a function of time for two different values of $\epsilon_0$ [pictured in (a)]. Dashed lines indicate the times of the LZSM transitions (i.e.\ $t$'s such that $\epsilon(t)=0$). For $\epsilon_0 = -39\ \mu$eV, the accumulated phases result in destructive interference and successive LZSM transitions cancel each other out (blue trace).  With $\epsilon_0 = -16.5\ \mu$eV (red trace) successive LZSM transitions interfere constructively, resulting in nearly complete transfer of population to the $\ket{1}$ state at $t$ $\approx$ 2 ns.}
\label{fig1}
\end{center}	
\vspace{-0.6cm}
\end{figure}

The LZSM problem describes the evolution of a TLS that is forced through an energy level anticrossing \cite{Landau,Zener,Stuckelberg,Majorana,ShevchenkoReview}. LZSM considered a generic TLS with states $\ket{0}$ and $\ket{1}$, described by the Hamiltonian:
\begin{equation}
H_{\rm TLS}=
\begin{pmatrix}
0 & \Delta  \\
\Delta & -\epsilon
\end{pmatrix}.
\label{Eq:TLS_H}
\end{equation}
Here the detuning parameter $\epsilon$ sets the energy difference between the states.  An off-diagonal matrix element $\Delta$ hybridizes the levels, resulting in an anticrossing of magnitude 2$\Delta$ at $\epsilon=0$ [Fig.~1(a)].

In the LZSM problem the energy difference between the states is varied with a linear level velocity $v = d | E_{\ket{1}} - E_{\ket{0}} | / dt$, where $E_{\ket{1}}$ ($E_{\ket{0}}$) is the energy of the $\ket{1}$ ($\ket{0}$) state.  This can be accomplished by driving the detuning according to $\epsilon(t) = v t$.  Starting in state $\ket{0}$ at time $t_i = -\infty$, the probability of remaining in state $\ket{0}$ at time $t_f = +\infty$ is given by the LZSM formula
\cite{Landau,Zener,Stuckelberg,Majorana}:
\begin{equation}
P_{\rm LZSM} = \exp \left(  -2 \pi \frac{\Delta^2}{\hbar v} \right),
\label{eq:LZ_p}
\end{equation}
where $\hbar$ is the reduced Planck's constant.  When $ \hbar v \ll \Delta^2$, $P_{\rm LZSM} \approx 0$ and the evolution is adiabatic. The system remains in the instantaneous eigenstate.  In the opposite limit $P_{\rm LZSM} \approx 1$ and the sudden change approximation can be made. Here a TLS that starts in $\ket{0}$ will remain in $\ket{0}$ after the sweep through the anticrossing.

With intermediate level velocities, a sweep through the anticrossing will generate a superposition of states $\ket{0}$ and $\ket{1}$ \cite{GefenPRL,SilliapaaPRL,PhysRevA.55.R2495}.  This physics has been harnessed for quantum control in a variety of systems, including Rydberg atoms \cite{PhysRevLett.68.3515,PhysRevLett.69.1919}, nitrogen vacancy centers \cite{NVlzs,ZhouPRL}, and GaAs DQDs \cite{PettaScience2010}.  

\subsection{Two-level dynamics under periodic driving}

The effects of quantum interference are revealed when the system is repeatedly driven through an anticrossing. Consider the case of sinusoidal driving.  Here $\epsilon(t) = \epsilon_0 + \epsilon_1 \sin\left(  2 \pi f t\right)$, where $\epsilon_0$ is a fixed detuning set by dc gate voltages in the experiment and $\epsilon_1$ = $e V_{\rm ac}$ is the amplitude of the ac drive. The ac drive results in two anticrossing traversals for each cycle of the drive field, with an approximate level velocity:
\begin{equation}
v \approx 2 \pi  \epsilon_1 f \sqrt{1-\left( \frac{\epsilon_0}{\epsilon_1}\right)^2}.
\label{eq:LZ_v}
\end{equation}
In Fig.~\ref{fig1}(b) we plot the energy levels as a function of time for sinusoidal driving with $f = 2$ GHz, $\epsilon_1 = 100\ \mu$eV, $\Delta = 4 \ \mu$eV, and $\epsilon_0 = -16.5 \ \mu$eV.  A system initialized in $\ket{0}$ at $t = 0$ will be forced through the anticrossing every time that $\epsilon_0 = - \epsilon_1 \sin \left( 2\pi f t\right)$. For our driving parameters the first crossing happens at $t = 0.01$ ns. After the first sweep through the anticrossing the probability of remaining in the $\ket{0}$ state is approximately $P_{\rm LZSM} = 0.9$. The non-unity probability results in the system entering a superposition of states $\ket{0}$ and $\ket{1}$. After the anticrossing the states accumulate a relative phase $\Phi_1$ due to their energy difference. At time $t=0.21$ ns the system is forced back through the anticrossing, interfering the two-paths of the interferometer.

Such interference occurs twice during each cycle and depends on the phases $\Phi_1$ and $\Phi_2$.  Additionally the total accumulated phase $\Phi_1+\Phi_2$ will result in interference between subsequent cycles.  Depending on the precise value of the phase the system will exhibit behavior ranging from constructive to destructive interference.  To illustrate this we plot the occupation of state $\ket{1}$, $P_{\ket{1}}$, as a function of time  in Fig.~\ref{fig1}(c).  We use two different values of the offset detuning $\epsilon_0$ and numerically integrate Schr{\"o}dinger's equation with the Hamiltonian in Eq.~\ref{Eq:TLS_H}. For $\epsilon_0 = - 16.5\ \mu$eV the phase accumulation results in constructive interference and $P_{\ket{1}}$ oscillates between 0 and 1.  For $\epsilon_0 = -39\ \mu$eV, however, the interference is destructive.  As such the population transfer resulting from one LZSM transition is immediately canceled out by the next LZSM transition.

In the fast driving limit (where $1 - P_{\rm LZSM}$ $\ll$ 1), the condition for constructive interference can be derived by considering the phase accumulation $\Phi_1$ occurring between the first and second LZSM transition and $\Phi_2$ occurring between the second and the third LZSM transition [see Fig.~\ref{fig1}(b)].  For constructive interference to occur $\Phi_1 - \Phi_2 = 2 n \pi$ for an integer $n$ \cite{ShevchenkoReview}.  With $\Phi_i = \int\limits_{t_i}^{t_{i+1}} \frac{E_{\ket{1}}-E_{\ket{0}}}{\hbar} dt \approx \int\limits_{t_i}^{t_{i+1}} \frac{\epsilon_0 + e V_{\rm ac} \sin\left( 2\pi f t \right)}{\hbar} dt$, the resonance condition reduces to $ n h f = \epsilon_0$.  Here $t_{i}$ is the time of the $i$-th anticrossing traversal.   This can be interpreted as an $n$-photon resonance condition and has been observed in several studies on both superconducting qubits \cite{OliverScience,BernNature} and GaAs charge qubits \cite{StehlikPRB}.

\label{sec:ThreeLevel}

\begin{figure}[t]
	\begin{center}
		\includegraphics[width=1\columnwidth]{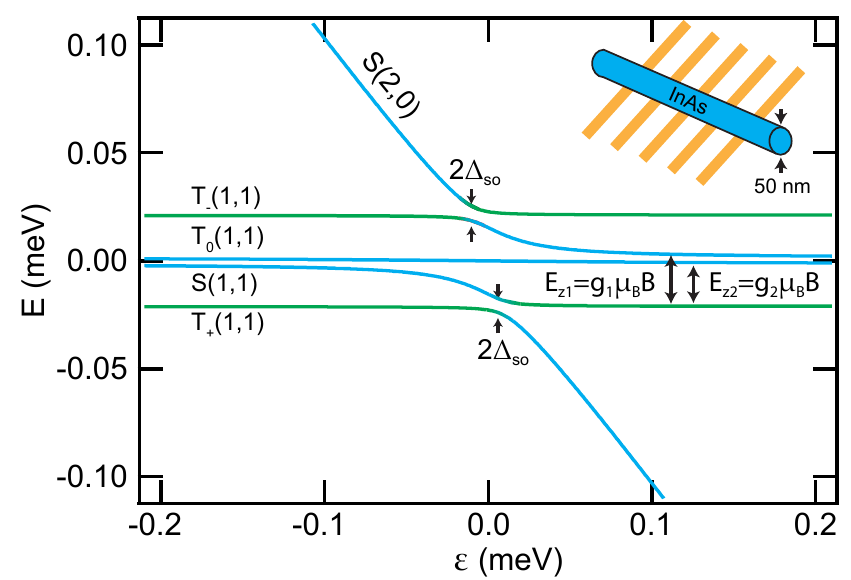}
		\caption[Double Quantum Dot Level Diagram]{\col DQD energy levels plotted as a function of detuning $\epsilon$ with $B = 45$ mT.  Interdot tunnel coupling $\Delta = 16.5\ \mu$eV hybridizes the S(2,0) and S(1,1) states, while the spin-orbit coupling $\Delta_{\rm so} = 4\ \mu$eV hybridizes the S(2,0) state with the  T$_+$(1,1) and T$_-$(1,1) states.  
		}
		\label{fig:harmonics:theory:model}
	\end{center}	
	\vspace{-0.6cm}
\end{figure}

\section{Periodically driven two-electron double quantum dot}
\label{sec:DQD}

EDSR experiments are typically performed near a Pauli-blocked interdot charge transition, where the total number of electrons in the DQD is even \cite{StefanNP,NowackEDSR}. We use a five-level Hamiltonian to capture the singlet-triplet physics of this system.  Starting with the DQD initialized in a spin-blocked triplet state, we show that multilevel LZSM interference leads to harmonics in the EDSR response near zero detuning \cite{RudnerPRL2014}.

\subsection{Double quantum dot energy level diagram}
To reflect the experimental conditions we consider a spin-orbit qubit defined in an InAs nanowire, as schematically shown in the inset of Fig.~\ref{fig:harmonics:theory:model}.  
We note that robust Pauli blockade has been observed in many experiments at higher electron occupancies \cite{ACJohnsonSpinBlocakde,SixElectron,StehlikPRL}.  For simplicity, we therefore consider a DQD in the two-electron regime, with one electron in each dot (1,1), or with two electrons in one dot, e.g.\ (2,0). Here we use the notation ($N_{\rm l}$,$N_{\rm r}$), where $N_{\rm l}$ ($N_{\rm r}$) is the number of electrons in the left (right) dot.  In the (1,1) charge configuration there are four spin states, the singlet state S(1,1) and the three triplet states T$_{-}$(1,1), T$_{0}$(1,1), T$_{+}$(1,1), with total spin components $m_{\rm s} = -1,0,+1$. An external magnetic field results in Zeeman splitting of the electronic spin states with  $E_{\rm z1} = g_1 \mu_{\rm B} B$ and $E_{\rm z2} = g_2 \mu_{\rm B}B$.  The g-factors are generally different due to strong spin-orbit coupling. We set $g_{1} = 7.8$ and $g_{2} = 6.8$  to match the values measured in Ref.~\cite{StehlikPRL}.
Due to the tight electric confinement there is a large singlet-triplet splitting $E_{\rm st} = 5.4$ meV.  As a result, the (2,0) triplet manifold can be neglected in most experimental situations. The Hamiltonian in the five-state basis ($\ket{S(1,1)}$, $\ket{S(2,0)}$, $\ket{T_0(1,1)}$, $\ket{T_-(1,1)}$, $\ket{T_+(1,1)}$) can be written as:
\begin{align}
& H_{\rm DQD}=  \nonumber \\
& \begin{pmatrix}
0		& \Delta 			&   \frac{E_{\rm z1}-E_{\rm z2}}{2} & 0 			 	& 0 \\
\Delta	& -\epsilon 		&	0								& \Delta_{\rm so}	& \Delta_{\rm so} \\
\frac{E_{\rm z1}-E_{\rm z2}}{2}& 0 & 0							& 0 				& 0 \\
0		& \Delta_{\rm so}	&	0								& \frac{E_{\rm z1}+E_{\rm z2}}{2} & 0 \\
0		& \Delta_{\rm so}	&	0								& 0						 & -\frac{E_{\rm z1}+E_{\rm z2}}{2}
\end{pmatrix}.
\label{Hdqd}
\end{align}
Here $\epsilon$ is the detuning, $\Delta$ is the interdot tunnel coupling, and $\Delta_{\rm so}$ generates spin-orbit anticrossings \cite{PhysRevB.77.045328,PergeHyperfine}.   The resulting energy level diagram is shown in Fig.~\ref{fig:harmonics:theory:model} with parameters $\Delta = 16.5\ \mu$eV, $\Delta_{\rm so} = 4\ \mu$eV \cite{koppensModeling,RevModPhys.75.1,KarlFancy,PergeHyperfine}.  These parameters are taken from Ref.~\cite{StehlikPRL} and the well-established material properties of InAs \cite{emass,epsilon,alpha110}.

\subsection{Time evolution of the five-level double quantum dot}

\begin{figure}
\begin{center}
\includegraphics[width=1\columnwidth]{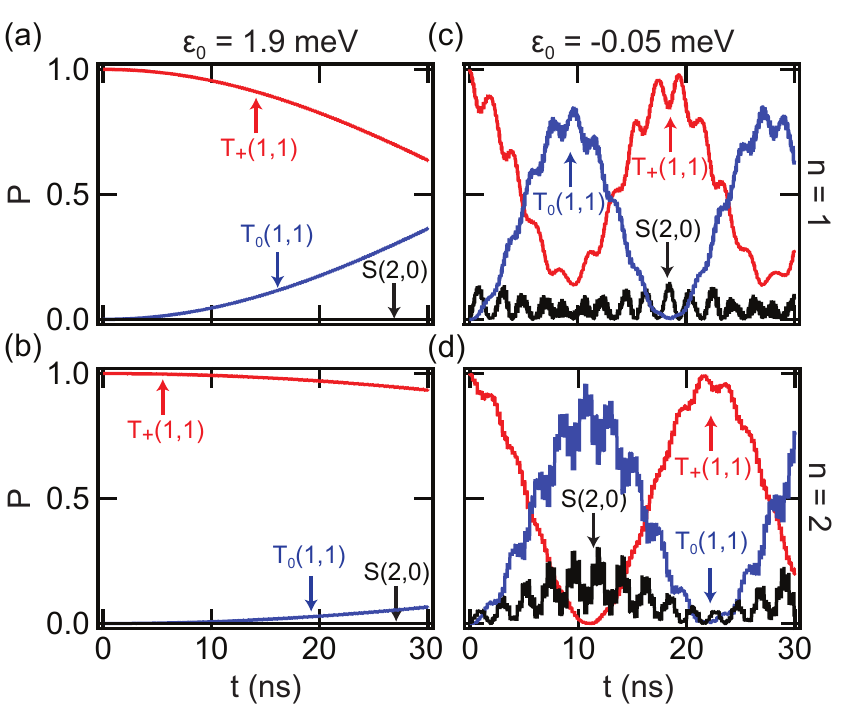}
\caption[Population Dynamics]{\col S(2,0), T$_{0}$(1,1), and T$_+$(1,1) occupation probabilities plotted as a function of time with $\epsilon_1 = 1.3$ meV and $B = 45$ mT.  The system is initialized in the T$_+$(1,1) state at $t = 0$. In (a,c) $f$ is chosen to drive a $n = 1$ photon EDSR process between the T$_+$(1,1) and T$_0$(1,1) states, while for (b,d)  $f$ is chosen to drive a $n = 2$ photon EDSR process between the  T$_+$(1,1) and T$_0$(1,1) states. There is a remarkable difference between the dynamics when the levels are far detuned [$\epsilon_0 = 1.9$ meV in (a,b)] compared to when the levels are near zero detuning [$\epsilon_0$ = -0.05 meV (c,d)]. Near zero detuning, spin transfer processes occur with significant population transfer into the S(2,0) state, and occur on a much faster timescale than in the case of far detuning. Since in all cases the S(1,1) and T$_-$(1,1) states (not shown) are not on resonance, there is no significant population transfer into these states.}
\label{fig:harmonics:theory:timetraces}
\end{center}
\end{figure}

To illustrate the importance of LZSM dynamics we time-evolve the five-level system described by Eq.~\ref{Hdqd}. This simple model reproduces the strong detuning dependence that is observed in the experimental data.  The system is initialized in the T$_{+}(1,1)$ state and propagated under an oscillatory detuning of the form $\epsilon(t) = \epsilon_0 + \epsilon_1 \sin \left( 2 \pi f t \right)$, with $B = 45$ mT, and $\epsilon_1 = 1.3$ meV. Figures~\ref{fig:harmonics:theory:timetraces}(a--d) show the T$_0$(1,1), T$_+$(1,1), and S(2,0) state occupations as a function of time. For panels (a,c) we choose driving that corresponds to a one-photon resonance between the T$_+$(1,1) and T$_0$(1,1) states.  For panels (b,d) the drive corresponds to a two-photon resonance between the T$_+$(1,1) and T$_0$(1,1) states. In the far detuned region ($\epsilon_0$ = 1.9 meV) the ac drive does not have a large enough amplitude to force the system through the anticrossings near $\epsilon$ = 0. In this case population transfer into the T$_{0}$(1,1) state is visible for the $n=1$ harmonic, as seen in Fig.~\ref{fig:harmonics:theory:timetraces}(a) \cite{GolovachLoss}. However, the dynamics for the $n=2$ resonance proceed on a significantly slower time scale, as expected from standard spin resonance theory \cite{cohen}. For both the $n=1$ and $n=2$ resonance conditions, there is no significant population transfer into the S(2,0) state [see Fig.~\ref{fig:harmonics:theory:timetraces}(a,b)].

When $\left|\epsilon_0\right| \lesssim \epsilon_1$ the dynamics are radically different.  Here the system is repeatedly forced through the level anticrossings, causing a portion of the population to be transferred to the S(2,0) state, from which the system can make further LZSM transitions to either the T$_+$(1,1) or T$_0$(1,1) state.  Evidence of these processes can be seen in Fig.~\ref{fig:harmonics:theory:timetraces}(c,d).  For both the $n$ = 1 and $n=2$ resonances clear population transfer is observed between the T$_{+}$(1,1) and T$_{\rm 0}$(1,1) states. The population transfer is mediated by the S(2,0) state, as evidenced by the periodic jumps of the S(2,0) state population. Since transitions to both the T$_- $(1,1) and S(1,1) states are not resonant, there is no significant population transfer into them.  Finally note that the timescale over which population transfer occurs is much shorter when $| \epsilon_0| \lesssim \epsilon_1$.

\begin{figure}
\begin{center}
\includegraphics[width=1\columnwidth]{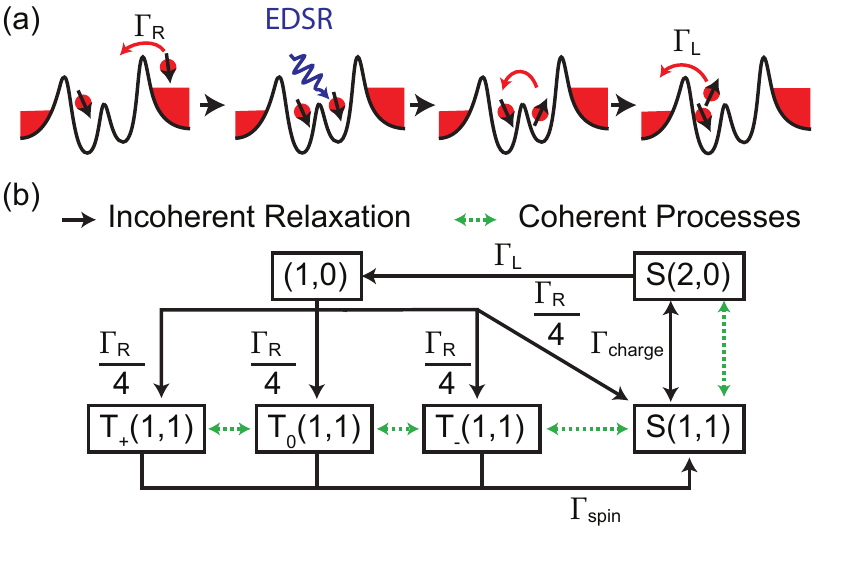}
\caption[Model of a Double Quantum Dot Transport Cycle]{\col (a)  Illustration of the charge transport cycle.  Starting from the (1,0) charge state an electron is loaded into the right dot.  If a (1,1) triplet state is loaded the transport becomes blocked until the spin is rotated into the S(1,1) state, from which the system can tunnel to S(2,0) and then into the left lead.   (b) Illustration of the various processes included in simulations of the transport dynamics. Black arrows indicate relaxation and incoherent tunneling processes, while green arrows indicate coherent processes that arise due to the periodic driving.} 
\label{fig:harmonics:theory:tunneling}
\end{center}
\end{figure}

\section{Transport cycle}
\label{sec:Results}

In the previously reported experiments \cite{StehlikPRL}, the EDSR response is detected by measuring the dc current through the DQD. The DQD is configured at finite bias in Pauli blockade. Resonant ac driving rotates the electronic spin states, lifting the Pauli blockade, resulting in a small, but measurable current \cite{StefanNP,NowackEDSR}.  To make a quantitative comparison with experiment we therefore model the full transport cycle of the DQD. As seen in Fig.~\ref{fig:harmonics:theory:tunneling}(a), starting from the empty (1,0) state, tunneling from the  right lead results in the (1,1) charge configuration.  If a polarized (1,1) triplet state is loaded, the transport cycle becomes blocked as tunneling into the (2,0) charge configuration is forbidden by the Pauli exclusion principle \cite{PauliBlockade}.  The Pauli blockade can be lifted by driving an EDSR transition, which rotates a blocked (1,1) state [T$_-$(1,1) or T$_+$(1,1)] to an unblocked state [S(1,1) or T$_0$(1,1)].  In our case T$_0$(1,1) is unblocked due to the difference in $g$-factors, which leads to further rotation to the S(1,1) state. Strong electron-phonon coupling results in fast relaxation from the S(1,1) state to the S(2,0) state. The electron then tunnels to the left lead with rate $\Gamma_{\rm L}$, completing the transport cycle.

\subsection{Time evolution}
We model the time dependence of the periodically driven system by evolving the density matrix $\rho$ using the master equation:

\begin{align}
\frac{d \rho}{dt} =& - \frac{i}{\hbar} \left[H_{\rm DQD} ,\rho \right] + \Gamma_{\rm L} \mathcal{D}\left[\ket{(1,0)}\bra{S(2,0)}  \right] \rho + \\
& \frac{\Gamma_{\rm R}}{4} \mathcal{D}\left[ \ket{S(1,1)}\bra{(1,0)}\right] \rho + \nonumber \\
& \sum_{j} \mathcal{D}\left[\ket{T_j(1,1)}\bra{(1,0)}\right] \rho + \nonumber \\
& \sum_{j} \Gamma_{\rm spin} \mathcal{D}\left[ \ket{S(1,1)}\bra{T_j(1,1)}\right] \rho + \nonumber \\
& \Gamma_{\rm S(2,0)} \mathcal{D} \left[ \ket{S(2,0)}\bra{S(2,0)} \right] \rho + \nonumber \\
& \Theta\left(\epsilon\right) \Gamma_{\rm charge} \mathcal{D}\left[\ket{S(2,0)}\bra{S(1,1)}\right] \rho + \nonumber \\
& \Theta\left(-\epsilon\right) \Gamma_{\rm charge} \mathcal{D}\left[\ket{S(1,1)}\bra{S(2,0)}\right] \rho  \nonumber.
\label{eq:Master}
\end{align}

\noindent Here $\mathcal{D}\left[A\right] \rho = -1/2 \{A^\dagger A,\rho \} +A\rho A^\dagger$  is the Lindblad superoperator describing relaxation and decoherence, $j$ spans the $m_{\rm s}$ = $0$, $+1$, and $-1$ triplet states. $\Theta(x)$ is the Heaviside step function.  As shown in Fig.~\ref{fig:harmonics:theory:tunneling}(b), terms with $\Gamma_{\rm L}$ ($\Gamma_{\rm R}$) account for coupling to the left (right) lead: the $\Gamma_{\rm L}$ term relaxes the S(2,0) state into the empty (1,0) state, while the $\Gamma_{\rm R}$ term moves the population from the empty state to one of the four (1,1) spin states.  Note that our model assumes unpolarized lead tunneling. Therefore the tunneling probability into any of the (1,1) spin states is equal.  $\Gamma_{\rm charge}$ is included to account for charge relaxation, which is known to take place on nanosecond timescales in semiconductor DQDs.  For $\epsilon > 0$, $\Gamma_{\rm charge}$ relaxes the S(1,1) state into the S(2,0) state, while for $\epsilon < 0$ this process is reversed.  This ensures that charge relaxation only takes place from a state of higher energy to a state of lower energy.  $\Gamma_{\rm spin}$ models spin relaxation, which is relatively slow in semiconductor DQDs. Finally $\Gamma_{\rm S(2,0)}$ [not pictured in Fig.~\ref{fig:harmonics:theory:tunneling}(b)] results in charge decoherence.  

\subsection{Charge transport and the role of decoherence}

\begin{figure}
\begin{center}
\includegraphics[width=1\columnwidth]{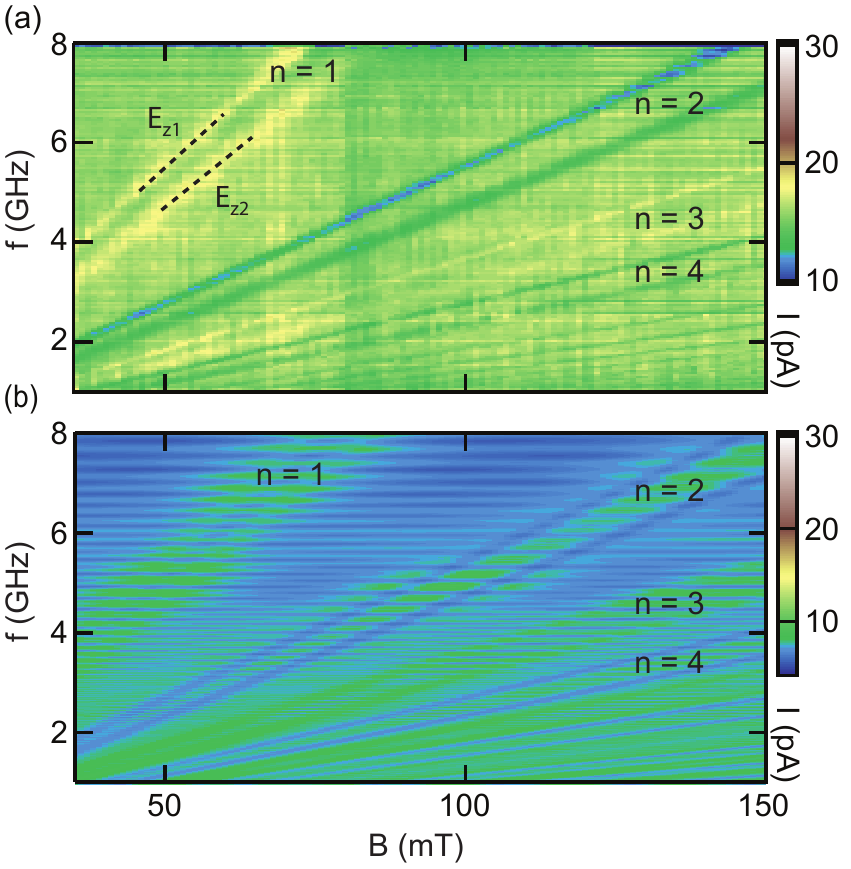}
\caption[Comparison Between Theory and Experiment]{\col (a) Measured current $I$ through the device as a function of magnetic field $B$ and frequency $f$. With the DQD configured at zero detuning a large number of multiphoton resonances are observed. Adapted from Ref.~\cite{StehlikPRL}.  (b) Calculated current through the DQD as a function of $B$ and $f$. The color-scale axes are slightly different since the off-resonance leakage current measured in experiment is device specific.}
\label{fig:harmonics:theory:Spectroscopy}
\end{center}
\end{figure}

We now simulate the experimental system using realistic parameters to account for tunnel coupling to the leads, charge noise, and dephasing. To match typical experimental conditions, we set $\Gamma_{\rm L} = \Gamma_{\rm R} = 2$ GHz,  $\Gamma_{\rm charge} = 1$ GHz, $\Gamma_{\rm S(2,0)} =10$ GHz, and $\Gamma_{\rm spin}=1$ MHz.  The parameters in the Hamiltonian describing the DQD are specified in Sec.~\ref{sec:DQD}.  For each value of $B$ and $f$ we initialize the system in the (1,0) state.  We then numerically propagate the system in time until the system reaches a steady state (typically after 20 ns of evolution). We can then write the current as $I = e \Gamma_{\rm R}  P_{(1,0)}$, where $P_{(1,0)}$ is the extracted steady state population of the (1,0) state.  We note that for typical drive parameters the minimum time between the T$_+$(1,1)$\leftrightarrow$S(2,0) and the S(2,0)$\leftrightarrow$T$_0$(1,1)  LZSM transitions is on the order of a picosecond.  As a result, the S(2,0) state can still act as an intermediary that transfers population between the other levels, despite the large relaxation and decoherence rates.

In Figs.~\ref{fig:harmonics:theory:Spectroscopy}(a,b) we compare the spectroscopic data obtained in the experiments with our model. Figure \ref{fig:harmonics:theory:Spectroscopy}(a) shows the data that were obtained with $\epsilon_0 =0$. Here the current $I$ is plotted as a function of magnetic field strength $B$ and the applied excitation frequency $f$.   For $n=1$ two distinct resonance lines of increased current are visible.  These correspond to $E_{\rm z1}$ ($g_1 = 7.8$) and $E_{\rm z2}$ ($g_2 = 6.8$).  Higher photon transitions display a striking odd-even dependence. Multiphoton resonances up to $n$ = 8 are observed. In Fig.~\ref{fig:harmonics:theory:Spectroscopy}(b) we plot the current $I$ as a function of $B$ and $f$, as calculated by the model described above. Following previous work, we include the effects of quasi-static charge noise by using Gaussian smoothing of the response \cite{KarlFancy,karlPRL,YinyuPRL}.  In this plot the effects of charge noise are included by sampling 30 different randomly chosen offset detunings and weighing the final response with a Gaussian of width $\sigma_{\rm charge} = 60\ \mu$eV centered around $\epsilon_0 = 0$.  The effects of the fluctuating nuclear field are included by smoothing the response in $B$ with a Gaussian of width 3.3 mT, which is the fluctuating Overhauser field measured in Ref.~\cite{StehlikPRL}. This Overhauser field is consistent with other values reported in the literature \cite{SchroerPRL}.

The results of our model replicate the overall structure of the experimental data.  Both the large number of higher photon transitions and the odd-even dependence of the leakage current are in qualitative agreement with the data. At a finer level, there are some slight deviations between the theoretical predictions and the experimental data. The observed current is in general higher then our model predicts.  We attribute this to imperfect fitting of the tunneling rates.  The simulations also exhibit faint high frequency oscillations (oriented horizontally in the figure) that are largely independent of $B$. We attribute this to the imperfect modeling of the charge noise. % Additional peaks and dips are also visible between the main resonances.  These correspond to a resonance between the T$_-$(1,1) and T$_+$(1,1) states, which have been recently theoretically investigated in Ref.~\cite{SanchesPRB}. %We attribute these deviations to imperfect modeling of all of the noise sources in the DQD. In particular, here we neglect the effects of nuclear field noise, as its full treatment is computationally prohibitive \cite{Ribeiro2013}.

\begin{figure}
\begin{center}
\includegraphics[width=1\columnwidth]{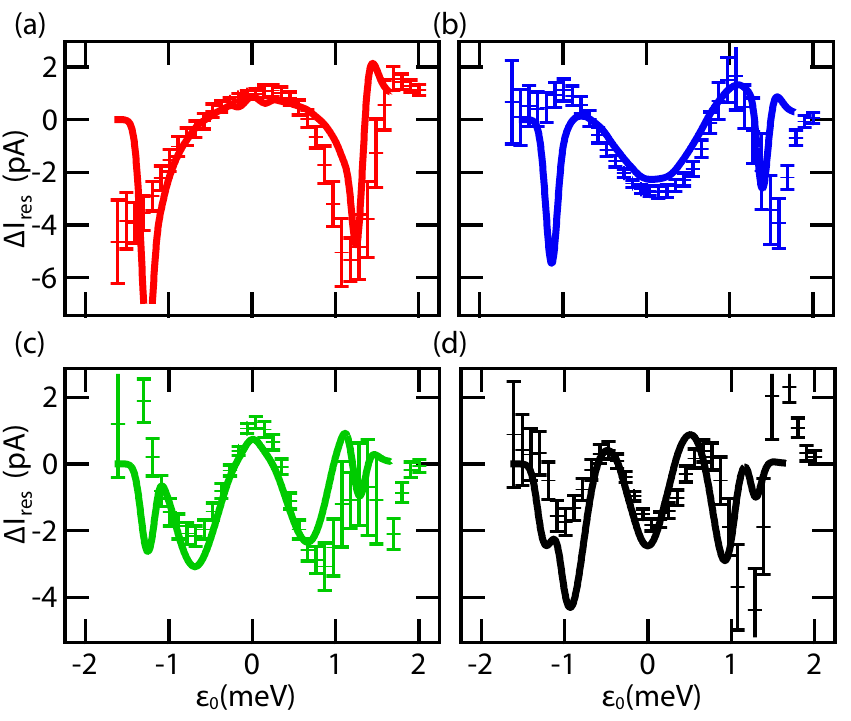}
\caption[Comparison of Detuning Dependence]{ \col Measured and simulated $\Delta I_{\rm res}$ (resonant change of current) as a function of $\epsilon_0$ for (a) $n=1$, (b) $n=2$, (c) $n=3$, (d) and $n=4$. }		\label{fig:harmonics:theory:DetuningDependence}	
\end{center}
\end{figure}

The experimental data also exhibit a characteristic detuning dependence. To make a valid comparison with experiments, we define $\Delta I_{\rm res}$ as the change between resonant and non-resonant leakage current.  In the experiment this quantity was obtained by measuring the current along $n hf = g_1 \mu_{\rm B} B$ and subtracting from it the current found approximately 5 mT away \cite{StehlikPRL}.  Figure \ref{fig:harmonics:theory:DetuningDependence} plots $\Delta I_{\rm res}$  for $\epsilon_1 = 1.3$ meV and $f = 4.7$ GHz (successive harmonics are achieved by increasing $B$).  The data points are adapted from Ref.~\cite{StehlikPRL}, while solid lines show the calculated $\Delta I_{\rm res}$ from the model.   % The response is highly oscillatory, with the oscillation period decreasing with $n$.  Additionally, while the even transitions have a local minimum at zero detuning, odd transitions have a local maximum.
%Solid lines in Fig.~\ref{fig:harmonics:theory:DetuningDependence} show the predicted $\Delta I_{\rm res}$ calculated from the model, for driving parameters  matching experimental conditions.  The numerical model is in good agreement with the measured data.  The overall magnitude of $\Delta I_{\rm res}$ matches very well.
We note that all experimentally observed features are reproduced.  First, the odd-even dependence is evident. Near zero detuning $\Delta I _{\rm res}$ has a maximum (minimum) for odd (even) photon resonances. Second, the number of oscillations in $\Delta I_{\rm res}$ increases with $n$, as observed in experiment. Lastly, the magnitude of $\Delta I_{\rm res}$ is in good agreement with the data. 
%$\Delta$ is in good agreement with the data. both the odd-even dependence and the increased number of oscillations with $n$ appear in the model.  We note a deviation in the width of some of the features, particularly the dip in $\Delta I_{\rm res}$ at $\epsilon_0 \approx e V_{\rm ac}$.  We attribute this discrepancy to the incomplete treatment of noise sources in our model.
%First, the even-odd dependence is evident. Near zero detuning \Delta ... is maximum (minimum) for odd (even)... Second, the number of oscillations in \Delta... increases with n, as observed in experiment. Lastly, the magnitude of 
%$\Delta$ is in good agreement with the data.

\section{Conclusion}

We have shown that the multiphoton resonances recently observed in EDSR experiments are due to multilevel LZSM interference.   The fact that these high order processes are possible raises several intriguing possibilities.  For example, since the mechanism for population transfer is quite distinct from traditional Rabi oscillations, one could obtain very fast population transfer, an attractive proposition for quantum manipulation \cite{PhysRevLett.110.066806,SachrajdaTQDLZS}.   With smaller charge noise it would also be possible to perform a direct measurement of the spin-orbit gaps \cite{shulman2014} and investigate the interplay of the spin-orbit and hyperfine interactions, both of which open gaps between the S(2,0) state and the T$_+$(1,1) and  T$_-$(1,1) states \cite{Quenching}.  Finally we note that so far experiments studying LZSM processes have focused on the zero detuning region near the singlet state anticrossing.  However, similar behavior should also be observable near anticrossings with the states in the (0,2) triplet manifold.  Due to the use of transport as a probe of spin states, these have so far been experimentally inaccessible.  However, recent developments of fast cavity based readout \cite{KarlFancy,PhysRevApplied.4.014018} should make this exciting regime within reach of experimental studies.

\begin{acknowledgements}
\vspace{-0.4cm}
We thank Sorawis Sangtawesin for assistance developing the simulation code. This research is funded by the Gordon and Betty Moore Foundation's EPiQS Initiative through Grant GBMF4535, with partial support from the National Science Foundation (DMR-1409556 and DMR-1420541). MZM and MHD acknowledge support from Fapesp and INCT-DISSE/CNPq, Brazil. Devices were fabricated in the Princeton University Quantum Device Nanofabrication Laboratory.
\end{acknowledgements}

\bibliography{MasterManaged}

\end{document}